\documentclass[twocolumn]{aastex61}
\submitjournal{ApJ}

\shorttitle{Discovery of the first low-luminosity quasar at $z > 7$ (SHELLQs VII)}
\shortauthors{Matsuoka et al.}

\begin{document}

\title{Discovery of the first low-luminosity quasar at {\scriptsize $z$} $> 7$}

\correspondingauthor{Yoshiki Matsuoka}
\email{yk.matsuoka@cosmos.ehime-u.ac.jp}

\author{Yoshiki Matsuoka}
\affil{Research Center for Space and Cosmic Evolution, Ehime University, Matsuyama, Ehime 790-8577, Japan.}

\author{Masafusa Onoue}
\affil{Max Planck Institut f\"{u}r Astronomie, K\"{o}nigstuhl 17, D-69117, Heidelberg, Germany}

\author{Nobunari Kashikawa}
\affil{Department of Astronomy, School of Science, The University of Tokyo, Tokyo 113-0033, Japan.}
\affil{National Astronomical Observatory of Japan, Mitaka, Tokyo 181-8588, Japan.}
\affil{Department of Astronomical Science, Graduate University for Advanced Studies (SOKENDAI), Mitaka, Tokyo 181-8588, Japan.}

\author{Michael A. Strauss}
\affil{Princeton University Observatory, Peyton Hall, Princeton, NJ 08544, USA.}

\author{Kazushi Iwasawa}
\affil{ICREA and Institut de Ci{\`e}ncies del Cosmos, Universitat de Barcelona, IEEC-UB, Mart{\'i} i Franqu{\`e}s, 1, 08028 Barcelona, Spain.}

\author{Chien-Hsiu Lee}
\affil{National Optical Astronomy Observatory, 950 North Cherry Avenue, Tucson, AZ 85719, USA.}

\author{Masatoshi Imanishi}
\affil{National Astronomical Observatory of Japan, Mitaka, Tokyo 181-8588, Japan.}
\affil{Department of Astronomical Science, Graduate University for Advanced Studies (SOKENDAI), Mitaka, Tokyo 181-8588, Japan.}

\author{Tohru Nagao}
\affil{Research Center for Space and Cosmic Evolution, Ehime University, Matsuyama, Ehime 790-8577, Japan.}

\author{Masayuki Akiyama}
\affil{Astronomical Institute, Tohoku University, Aoba, Sendai, 980-8578, Japan.}

\author{Naoko Asami}
\affil{Seisa University, Hakone-machi, Kanagawa, 250-0631, Japan.}

\author{James Bosch}
\affil{Princeton University Observatory, Peyton Hall, Princeton, NJ 08544, USA.}

\author{Hisanori Furusawa}
\affil{National Astronomical Observatory of Japan, Mitaka, Tokyo 181-8588, Japan.}

\author{Tomotsugu Goto}
\affil{Institute of Astronomy and Department of Physics, National Tsing Hua University, Hsinchu 30013, Taiwan.}

\author{James E. Gunn}
\affil{Princeton University Observatory, Peyton Hall, Princeton, NJ 08544, USA.}

\author{Yuichi Harikane}
\affil{Institute for Cosmic Ray Research, The University of Tokyo, Kashiwa, Chiba 277-8582, Japan}
\affil{Department of Physics, Graduate School of Science, The University of Tokyo, Bunkyo, Tokyo 113-0033, Japan}

\author{Hiroyuki Ikeda}
\affil{National Astronomical Observatory of Japan, Mitaka, Tokyo 181-8588, Japan.}

\author{Takuma Izumi}
\affil{National Astronomical Observatory of Japan, Mitaka, Tokyo 181-8588, Japan.}

\author{Toshihiro Kawaguchi}
\affil{Department of Economics, Management and Information Science, Onomichi City University, Onomichi, Hiroshima 722-8506, Japan.}

\author{Nanako Kato}
\affil{Graduate School of Science and Engineering, Ehime University, Matsuyama, Ehime 790-8577, Japan.}

\author{Satoshi Kikuta}
\affil{National Astronomical Observatory of Japan, Mitaka, Tokyo 181-8588, Japan.}
\affil{Department of Astronomical Science, Graduate University for Advanced Studies (SOKENDAI), Mitaka, Tokyo 181-8588, Japan.}

\author{Kotaro Kohno}
\affil{Institute of Astronomy, The University of Tokyo, Mitaka, Tokyo 181-0015, Japan.}
\affil{Research Center for the Early Universe, University of Tokyo, Tokyo 113-0033, Japan.}

\author{Yutaka Komiyama}
\affil{National Astronomical Observatory of Japan, Mitaka, Tokyo 181-8588, Japan.}
\affil{Department of Astronomical Science, Graduate University for Advanced Studies (SOKENDAI), Mitaka, Tokyo 181-8588, Japan.}

\author{Shuhei Koyama}
\affil{Research Center for Space and Cosmic Evolution, Ehime University, Matsuyama, Ehime 790-8577, Japan.}

\author{Robert H. Lupton}
\affil{Princeton University Observatory, Peyton Hall, Princeton, NJ 08544, USA.}

\author{Takeo Minezaki}
\affil{Institute of Astronomy, The University of Tokyo, Mitaka, Tokyo 181-0015, Japan.}

\author{Satoshi Miyazaki}
\affil{National Astronomical Observatory of Japan, Mitaka, Tokyo 181-8588, Japan.}
\affil{Department of Astronomical Science, Graduate University for Advanced Studies (SOKENDAI), Mitaka, Tokyo 181-8588, Japan.}


\author{Hitoshi Murayama}
\affil{Kavli Institute for the Physics and Mathematics of the Universe, WPI, The University of Tokyo,Kashiwa, Chiba 277-8583, Japan.}

\author{Mana Niida}
\affil{Graduate School of Science and Engineering, Ehime University, Matsuyama, Ehime 790-8577, Japan.}

\author{Atsushi J. Nishizawa}
\affil{Institute for Advanced Research, Nagoya University, Furo-cho, Chikusa-ku, Nagoya 464-8602, Japan.}

\author{Akatoki Noboriguchi}
\affil{Graduate School of Science and Engineering, Ehime University, Matsuyama, Ehime 790-8577, Japan.}

\author{Masamune Oguri}
\affil{Department of Physics, Graduate School of Science, The University of Tokyo, Bunkyo, Tokyo 113-0033, Japan}
\affil{Kavli Institute for the Physics and Mathematics of the Universe, WPI, The University of Tokyo,Kashiwa, Chiba 277-8583, Japan.}
\affil{Research Center for the Early Universe, University of Tokyo, Tokyo 113-0033, Japan.}

\author{Yoshiaki Ono}
\affil{Institute for Cosmic Ray Research, The University of Tokyo, Kashiwa, Chiba 277-8582, Japan}

\author{Masami Ouchi}
\affil{Institute for Cosmic Ray Research, The University of Tokyo, Kashiwa, Chiba 277-8582, Japan}
\affil{Kavli Institute for the Physics and Mathematics of the Universe, WPI, The University of Tokyo,Kashiwa, Chiba 277-8583, Japan.}

\author{Paul A. Price}
\affil{Princeton University Observatory, Peyton Hall, Princeton, NJ 08544, USA.}

\author{Hiroaki Sameshima}
\affil{Koyama Astronomical Observatory, Kyoto-Sangyo University, Kita, Kyoto, 603-8555, Japan.}

\author{Andreas Schulze}
\affil{National Astronomical Observatory of Japan, Mitaka, Tokyo 181-8588, Japan.}

\author{Hikari Shirakata}
\affil{Department of Cosmosciences, Graduates School of Science, Hokkaido University, N10 W8, Kitaku, Sapporo 060-0810, Japan.}

\author{John D. Silverman}
\affil{Kavli Institute for the Physics and Mathematics of the Universe, WPI, The University of Tokyo,Kashiwa, Chiba 277-8583, Japan.}

\author{Naoshi Sugiyama}
\affil{Kavli Institute for the Physics and Mathematics of the Universe, WPI, The University of Tokyo,Kashiwa, Chiba 277-8583, Japan.}
\affil{Graduate School of Science, Nagoya University, Furo-cho, Chikusa-ku, Nagoya 464-8602, Japan.}

\author{Philip J. Tait}
\affil{Subaru Telescope, Hilo, HI 96720, USA.}

\author{Masahiro Takada}
\affil{Kavli Institute for the Physics and Mathematics of the Universe, WPI, The University of Tokyo,Kashiwa, Chiba 277-8583, Japan.}

\author{Tadafumi Takata}
\affil{National Astronomical Observatory of Japan, Mitaka, Tokyo 181-8588, Japan.}
\affil{Department of Astronomical Science, Graduate University for Advanced Studies (SOKENDAI), Mitaka, Tokyo 181-8588, Japan.}

\author{Masayuki Tanaka}
\affil{National Astronomical Observatory of Japan, Mitaka, Tokyo 181-8588, Japan.}
\affil{Department of Astronomical Science, Graduate University for Advanced Studies (SOKENDAI), Mitaka, Tokyo 181-8588, Japan.}

\author{Ji-Jia Tang}
\affil{Institute of Astronomy and Astrophysics, Academia Sinica, Taipei, 10617, Taiwan.}

\author{Yoshiki Toba}
\affil{Department of Astronomy, Kyoto University, Sakyo-ku, Kyoto, Kyoto 606-8502, Japan.}
\affil{Institute of Astronomy and Astrophysics, Academia Sinica, Taipei, 10617, Taiwan.}

\author{Yousuke Utsumi}
\affil{Kavli Institute for Particle Astrophysics and Cosmology, Stanford University, CA 94025, USA.}

\author{Shiang-Yu Wang}
\affil{Institute of Astronomy and Astrophysics, Academia Sinica, Taipei, 10617, Taiwan.}

\author{Takuji Yamashita}
\affil{Research Center for Space and Cosmic Evolution, Ehime University, Matsuyama, Ehime 790-8577, Japan.}



\begin{abstract}
We report the discovery of a quasar at $z = 7.07$, which was selected from the deep multi-band imaging data collected by the Hyper Suprime-Cam (HSC) Subaru Strategic Program survey.
This quasar, HSC $J124353.93+010038.5$, has an order of magnitude lower luminosity than do the other known quasars at $z > 7$.
The rest-frame ultraviolet absolute magnitude is $M_{1450} = -24.13 \pm 0.08$ mag and the bolometric luminosity is $L_{\rm bol} = (1.4 \pm 0.1) \times 10^{46}$ erg s$^{-1}$.
Its spectrum in the optical to near-infrared shows strong emission lines, and shows evidence for a fast gas outflow, as the \ion{C}{4} line is blueshifted and there is indication of broad absorption lines.
The \ion{Mg}{2}-based black hole mass is $M_{\rm BH} = (3.3 \pm 2.0) \times 10^8 M_\odot$, thus indicating a moderate mass accretion rate with an Eddington ratio $\lambda_{\rm Edd} = 0.34 \pm 0.20$.
It is the first $z > 7$ quasar with sub-Eddington accretion, besides being the third most distant quasar, known to date.
The luminosity and black hole mass are comparable to, or even lower than, those measured for the majority of low-$z$ quasars discovered by the Sloan Digital Sky Survey,
and thus this quasar likely represents a $z > 7$ counterpart to quasars commonly observed in the low-$z$ universe.

\end{abstract}




\section{Introduction} \label{sec:intro}

Quasars residing in the first billion years of the Universe ($z > 5.7$) have been used as various types of probes into early cosmic history.
The progress of cosmic reionization can be estimated from \ion{H}{1} absorption imprinted on the rest-frame ultraviolet spectrum of a high-$z$ quasar;  
this absorption is very sensitive to the neutral fraction of the foreground intergalactic medium \citep[IGM;][]{gunn65, fan06araa}.
The luminosity and mass functions of quasars reflect the seeding and growth mechanisms of supermassive black holes (SMBHs), which can be studied through comparison with theoretical models
\citep[e.g.,][]{volonteri12, ferrara14, madau14}.
Measurements of quasar host galaxies and surrounding environments tell us about the earliest mass assembly, possibly happening in the highest density peaks of the underlying
dark matter distribution \citep[e.g.,][]{goto09, decarli17, izumi18}.

Quasars at the highest redshifts are of particular interest,
as they have spent only a short time since their formation.
The current frontier for high-$z$ quasar searches is $z > 7$, where only a few quasars have been found to date.
Since radiation from $z > 7$ quasars is almost completely absorbed by the IGM at observed wavelengths $\lambda <$ 9700 \AA, and such objects are very rare and faint, one needs wide-field deep imaging
in near-infrared (IR) bands or in the $y$-band with red-sensitive CCDs to discover those quasars.
The first $z > 7$ quasar was discovered by \citet{mortlock11} at $z = 7.09$, from the United Kingdom Infrared Telescope (UKIRT) Infrared Deep Sky Survey \citep[UKIDSS;][]{lawrence07} data.
The second one was discovered by \citet{banados18} at $z = 7.54$, by combining data from the {\it Wide-field Infrared Survey Explorer} \citep{wright10}, UKIDSS, and the Dark Energy Camera Legacy Survey\footnote{
http://legacysurvey.org/decamls}.
In addition, two quasars, both at $z = 7.02$, were recently discovered \citep{wang18, yang18} by combining datasets from several wide-field surveys, including the Dark Energy Survey \citep{des05},
the Dark Energy Spectroscopic Instrument legacy imaging surveys \citep{dey18}, and
the Panoramic Survey Telescope \& Rapid Response System 1 \citep[Pan-STARRS1;][]{chambers16}.

However, the above $z > 7$ quasars are all very luminous \citep[if they are not strongly lensed; e.g.,][]{fan18, pacucci18}, due to the detection limits of the imaging survey observations.
These quasars harbor SMBHs with masses of roughly a billion solar masses, shining at close to the Eddington luminosity 
\citep[however the black hole mass of one of the quasars at $z = 7.02$ has not been measured;][]{yang18}.
They likely represent the most extreme monsters, which are very rare at all redshifts, especially at $z > 7$.
To understand a wider picture of the formation and early evolution of SMBHs, it is crucial to find $z > 7$ quasars of more typical luminosity, which would be direct counterparts to low-$z$ ordinary quasars.

This letter presents the discovery of a quasar at $z = 7.07$, HSC $J124353.93+010038.5$ (hereafter ``$J1243+0100$"), which has an order of magnitude lower luminosity than do the other known $z > 7$ quasars.
It harbors a SMBH with a mass of $M_{\rm BH} = (3.3 \pm 2.0) \times 10^8 M_\odot$ and shining at an Eddington ratio $\lambda_{\rm Edd} = 0.34 \pm 0.20$.
We describe the target selection and spectroscopic observations in \S \ref{sec:obs}.
The spectral properties of the quasar are measured and discussed in \S \ref{sec:results}.
A summary appears in \S \ref{sec:summary}.
We adopt the cosmological parameters $H_0$ = 70 km s$^{-1}$ Mpc$^{-1}$, $\Omega_{\rm M}$ = 0.3, and $\Omega_{\rm \Lambda}$ = 0.7.
All magnitudes refer to point spread function (PSF) magnitudes in the AB system \citep{oke83}, and are corrected for Galactic extinction \citep{schlegel98}.

\section{Observations} \label{sec:obs}

\begin{figure}
\epsscale{1.0}
\plotone{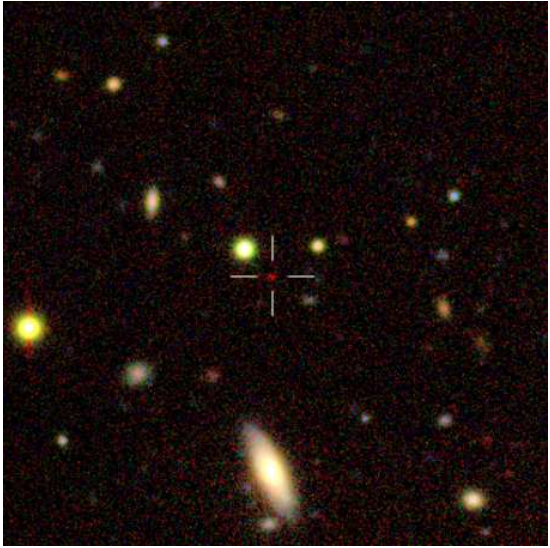}
\caption{Three-color (HSC $i$, $z$, and $y$-band) composite image around $J1243+0100$, marked with the cross-hair.
The image size is 1 arcmin  on a side. North is up and East to the left.
\label{fig:images}}
\end{figure}

$J1243+0100$ was selected from the Hyper Suprime-Cam (HSC) Subaru Strategic Program (SSP) survey \citep{aihara17_survey} data,
as a part of the ``Subaru High-$z$ Exploration of Low-Luminosity Quasars (SHELLQs)'' project \citep{matsuoka16, matsuoka18a,matsuoka18b,matsuoka18c}.
The coordinates and brightness are summarized in Table \ref{tab:brightness}.
A three-color composite image around the quasar is 
presented in Figure \ref{fig:images}.
This source has a full-width-at-half-maximum (FWHM) size of 0\arcsec.7 on the $y$-band image, which is consistent with the PSF size estimated at the corresponding
image position.
We used the methods detailed in \citet{matsuoka18b} to select this source as a high-$z$ quasar candidate.
The probability that this source was a quasar, not a Galactic brown dwarf, was $P_{\rm Q} = 0.4$, based on our Bayesian probabilistic algorithm \citep{matsuoka16} 
and the HSC $i$, $z$, and $y$-band photometry. 
It is among $\sim$30 $z$-band dropout objects in our quasar candidate list; we have so far conducted follow-up observations of roughly half of these candidates, and
partly reported the results in the SHELLQs papers mentioned above.
The highest-$z$ quasar we found and published previously is at $z \sim 6.9$ \citep{matsuoka18a}.

\begin{table}[h!]
\centering
\caption{Coordinates\footnote{Coordinates are at $J$2000.0. The astrometric accuracy of the HSC-SSP data is estimated to be 0\arcsec.1 \citep{aihara17_pdr1}.} and Brightness} \label{tab:brightness}
\begin{tabular}{cD@{$\pm$}D}
\tablewidth{0pt}
\hline
R.A. & 12$^{\rm h}$43$^{\rm m}$53$^{\rm s}$.93\\
Decl. & +01$^{\circ}$00\arcmin38\arcsec.5\\
$g_{\rm AB}$ (mag) & $<$26.7 (2$\sigma$)\\
$r_{\rm AB}$ (mag) & $<$26.5 (2$\sigma$)\\
$i_{\rm AB}$ (mag) & $<$26.7 (2$\sigma$)\\
$z_{\rm AB}$ (mag) & $<$25.8 (2$\sigma$)\\
$y_{\rm AB}$ (mag) & 23.57 $\pm$ 0.08\\
$m_{1450}$ (mag) & 22.82 $\pm$ 0.08\\
$M_{1450}$ (mag) & $-24.13 \pm 0.08$\\
$L_{\rm bol}$ (erg s$^{-1}$) & (1.4 $\pm$ 0.1) $\times 10^{46}$\\
\hline
\end{tabular}
\end{table}

We obtained a red-optical spectrum of the candidate using the Faint Object Camera and Spectrograph \citep[FOCAS;][]{kashikawa02} mounted on the Subaru telescope.
The observations were carried out on 2018 April 24 as a part of a Subaru intensive program (program ID: S16B-011I). 
We used FOCAS in the multi-object spectrograph mode with the VPH900 grism and SO58 order-sorting filter.
With a slit width of 1\arcsec.0, this configuration gave spectral coverage from 0.75 to 1.05 $\mu$m and resolution $R \sim 1200$.
We took 7 $\times$ 10-min exposures with 1\arcsec\ dithering between exposures along the slit, under photometric skys with the seeing around 0\arcsec.6.
The data were reduced with IRAF using the dedicated FOCASRED package in the standard manner.
The wavelength scale was calibrated with reference to sky emission lines, and the flux calibration was tied to a white dwarf standard star, Feige 34, observed on the same night.
The slit loss was corrected for by scaling the spectrum to match the HSC $y$-band magnitude (Table \ref{tab:brightness}).

A near-IR spectrum of the object was obtained with the Gemini Near-InfraRed Spectrograph \citep[GNIRS;][]{elias06} on the Gemini-north telescope.
The observations were carried out on 2018 June 25, July 22 and 29 in the queue mode (program ID: GN-2018A-FT-112).
We used the cross-dispersed mode with 32 l/mm grating, with the central wavelength set to 1.65 $\mu$m.
The slit width was 1\arcsec.0, which gave spectral coverage from 0.85 to 2.5 $\mu$m and resolution $R \sim 500$.
We took 36 $\times$ 5-min exposures in total, with 3\arcsec\ dithering between exposures along the slit, under spectroscopic skys with the seeing 0\arcsec.5 -- 0\arcsec.7.
The data reduction was performed with IRAF using the Gemini GNIRS package, in the standard manner.
The wavelength scale was calibrated with reference to Argon lamp spectra, and the flux calibration and telluric absorption correction were tied to a standard star, HIP 58510, 
observed right before or after the quasar observations at similar airmass.
We scaled the GNIRS spectrum to match the FOCAS spectrum in the overlapping wavelength range. 

In addition, we took a $K$-band spectrum of the quasar with the Multi-Object Infrared Camera and Spectrograph \citep[MOIRCS;][]{ichikawa06} on the Subaru telescope.
The observations were carried out on 2018 July 8 and 9 (program ID: S18A-061).
We used MOIRCS in the multi-object spectrograph mode with the VPH-$K$ grism.
The slit width was 0\arcsec.8, which gave spectral coverage from 1.8 to 2.5 $\mu$m and resolution $R \sim 1700$.
We took 34 $\times$ 4-min exposures in total, with 3\arcsec\ dithering between exposures along the slit, under spectroscopic skys with the seeing 0\arcsec.5 -- 0\arcsec.8.
The data reduction was performed with IRAF using the MCSMDP package, in the standard manner.
The wavelength scale was calibrated with reference to sky emission lines, and the flux calibration and telluric absorption correction were tied to a standard star, HIP 69747, 
observed right after the quasar observations.
We scaled the MOIRCS spectrum to match the GNIRS spectrum in the overlapping wavelength range. 

\begin{figure*}
\epsscale{1.1}
\plotone{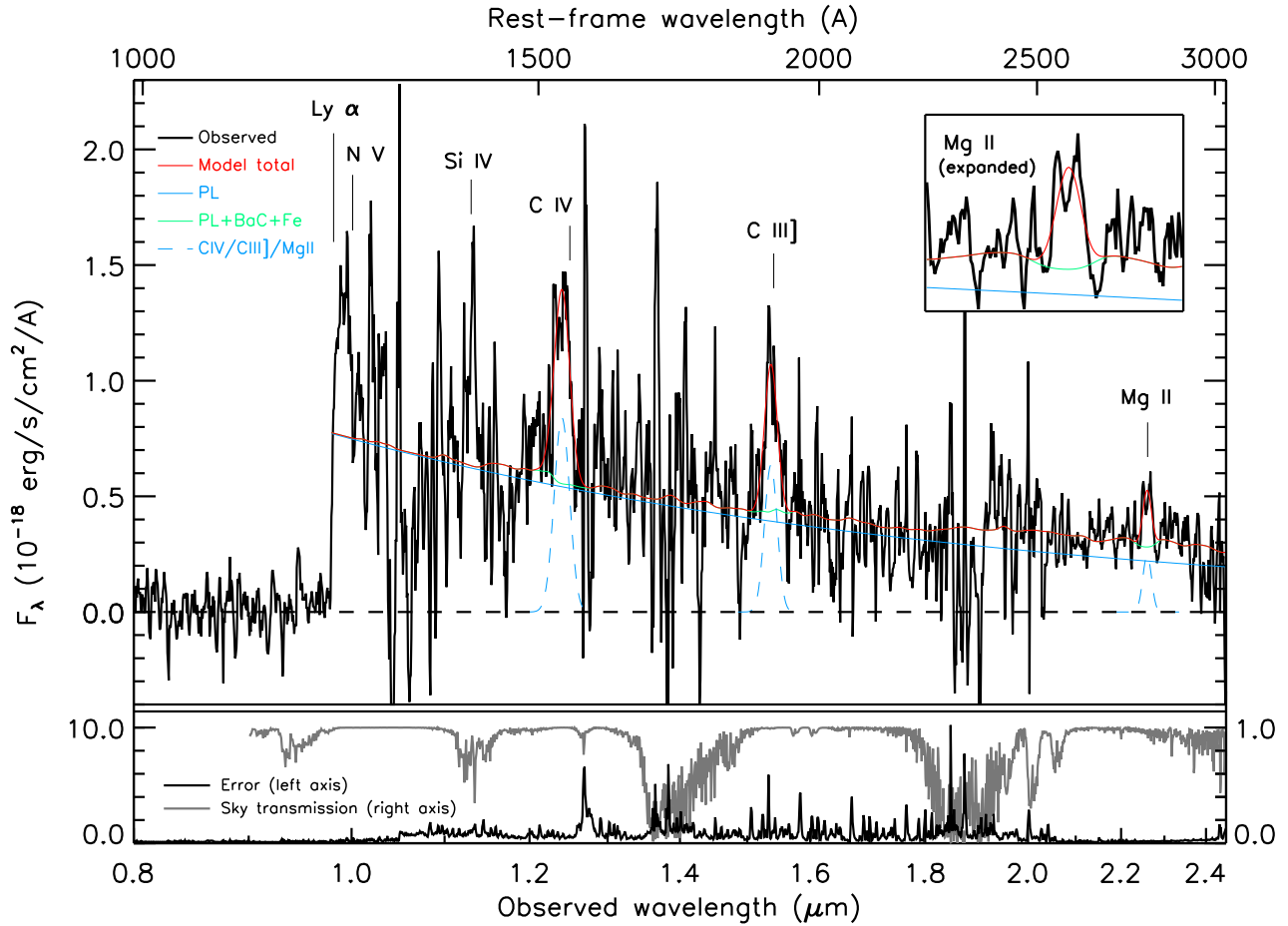}
\caption{Upper panel: the merged FOCAS, GNIRS, and MOIRCS spectrum of $J1243+0100$, smoothed using inverse-variance weighted means in 5 pixel for display purpose (black line). 
The expected positions of strong emission lines are marked by the vertical lines, given the \ion{Mg}{2}-based redshift.
Also plotted is the best-fit model spectrum (red line), which is the sum of a power-law continuum (blue solid line), Balmer and iron continua (the green solid line represents the sum of these two components plus the power-law continuum), the emission lines of \ion{C}{4} $\lambda$1549, \ion{C}{3}] $\lambda$1909, and \ion{Mg}{2} $\lambda$2800 (blue dashed lines).
The dips of observed flux relative to the power-law continuum model at $\sim$1.05 $\mu$m and $\sim$1.17 $\mu$m may be due to BALs.
The insert shows an expanded view of the spectrum around \ion{Mg}{2}.
Lower panel: the error spectrum (black line) and the atmospheric transmission spectrum above Maunakea \citep[][data retrieved from Gemini Observatory and plotted in gray]{lord92}.
\label{fig:spec}}
\end{figure*}

Finally, the FOCAS, GNIRS, and MOIRCS data were merged into a single spectrum, with a wavelength pixel spacing of $\lambda$/$\Delta\lambda$ = 1500. 
The associated errors were derived from the sky background spectrum measured for each of the above observations, and were propagated to the final spectrum. 
Figure \ref{fig:spec} presents the merged spectrum and errors, which are used for the measurements described in the following section.
While the spectrum may show marginally positive flux in the \citet{gunn65} trough bluewards of Ly$\alpha$, these are likely due to imperfect sky subtraction, 
as we see no signal at $<$0.98 $\mu$m in the 2d spectrum.
The \ion{Mg}{2} line appears to have two peaks, but these peaks do not appear consistently among the individual exposures and are likely due to noise.

\section{Spectral Measurements} \label{sec:results}

We measured spectral properties of the quasar through model fitting. 
The model consists of a power-law continuum with a slope $\alpha_\lambda = -1.5$ \citep{vandenberk01}\footnote{
We also tried fitting with a variable slope and found that this parameter is poorly constrained ($\alpha_\lambda = -2.0 \pm 0.8$), presumably due to degeneracy with
the other continuum components and the limited signal-to-noise ratio (S/N) of the present data.
}, 
Balmer continuum \citep{derosa11}, \ion{Fe}{2} $+$ \ion{Fe}{3} pseudo-continuum 
\citep[][modified following \citet{kurk07}]{vestergaard01, tsuzuki06}, and Gaussian profiles to represent the \ion{C}{4} $\lambda$1549, \ion{C}{3}] $\lambda$1909, and \ion{Mg}{2} $\lambda$2800 emission lines.
Each emission line was modeled with a single Gaussian, given the limited S/N of the spectrum.
The fitting was performed in the rest-frame wavelength range from $\lambda_{\rm rest}$ = 1450 \AA\ to 3000 \AA, which contains the three emission lines listed above\footnote{
We did not include in the fitting the spectral region around \ion{Si}{4} $\lambda$1400, which is affected by the low atmospheric transmission and possibly by broad absorption lines (BALs; see below).
}. 
All model components were fitted simultaneously to the data via $\chi^2$ minimization, which provided the best-fit parameter values and associated errors.
The derived best-fit model is presented in Figure \ref{fig:spec}.

We measured the apparent and absolute magnitudes of the quasar at $\lambda_{\rm rest}$ = 1450 \AA\ from the best-fit power-law continuum. 
We also measured the continuum luminosity at $\lambda_{\rm rest}$ = 3000 \AA\ from the best-fit model, and converted to the bolometric luminosity assuming a bolometric correction factor
BC$_{3000}$ = 5.15 \citep{shen11}.
The results are listed in Table \ref{tab:brightness}.

Table \ref{tab:spec} summarizes the emission line properties derived from the best-fit Gaussian models.
The quasar redshift measured from the \ion{Mg}{2} line is $z = 7.07 \pm 0.01$.
We found that the emission lines from the higher ionization species, \ion{C}{4} in particular, are significantly blueshifted relative to \ion{Mg}{2}.
We estimated the black hole mass ($M_{\rm BH}$) from the FWHM of \ion{Mg}{2} and the continuum luminosity at $\lambda_{\rm rest}$ = 1350 \AA, 
using the calibration presented by \citet{vestergaard09}
\footnote{We could use the continuum luminosity estimated at a wavelength closer to the \ion{Mg}{2} line, but given the limited S/N of the present data, the measurement would be
affected by the degeneracy of the three (power-law, Balmer, and iron) continuum components.
We checked that the $M_{\rm BH}$ estimate doesn't change significantly when the continuum luminosity at 2100 or 3000 \AA\ is used alternatively, with the corresponding calibration factor from \citet{vestergaard09}.}.
This yielded $M_{\rm BH} =  (3.3 \pm 2.0) \times 10^8 M_\odot$, and an Eddington ratio $\lambda_{\rm Edd} = 0.34 \pm 0.20$.
The systematic uncertainty of the above calibration is estimated to be a factor of a few, which is not included in the $M_{\rm BH}$ and  $\lambda_{\rm Edd}$ errors presented in this letter.

The Ly $\alpha$ strength relative to the above emission lines appears weaker than observed in low-$z$ quasars \citep[e.g.,][]{vandenberk01}.
This is likely due to IGM absorption, including damping wing absorption redwards of Ly $\alpha$, and/or possible BALs.
We defer detailed modeling of these absorptions to a future paper, and here measured the Ly $\alpha$ $+$ \ion{N}{5} $\lambda$1240 flux by simply integrating observed flux excess 
above the continuum model over $\lambda_{\rm rest} = 1215 - 1255$ \AA; the result is listed in Table \ref{tab:spec}.

\begin{deluxetable*}{rcccc}
\tablecaption{Emission line measurements \label{tab:spec}}
\tabletypesize{\scriptsize}
\tablehead{
\colhead{} &  \colhead{Ly $\alpha$ + \ion{N}{5} $\lambda$1240} & \colhead{\ion{C}{4} $\lambda$1549} & \colhead{\ion{C}{3}] $\lambda$1909} & \colhead{\ion{Mg}{2} $\lambda$2800}
} 
\startdata
Redshift                                    & \nodata & \nodata                                               & \nodata                                               & 7.07 $\pm$ 0.01\\
Velocity offset (km s$^{-1}$)      & \nodata       & $-$2400 $\pm$ 500                               &  $-$800 $\pm$ 400                             & \nodata  \\
Flux (erg s$^{-1}$ cm$^{-2}$)  & (9.6 $\pm$ 0.4) $\times$ 10$^{-17}$ & (2.1 $\pm$ 0.4) $\times$ 10$^{-16}$   & (1.6 $\pm$ 0.5) $\times$ 10$^{-16}$ & (6.2 $\pm$ 1.9) $\times$ 10$^{-17}$ \\
REW (\AA)                            & 16 $\pm$ 1 &  48 $\pm$ 10                                    &  51 $\pm$ 15                                 & 35 $\pm$ 11  \\
FWHM (km s$^{-1}$)               & \nodata & 5500 $\pm$ 1300                             &  4600 $\pm$ 1500                         & 3100 $\pm$ 900  \\
$M_{\rm BH}$ ($M_\odot$)     & \nodata & \nodata                                             & \nodata                                          & (3.3 $\pm$ 2.0) $\times$ 10$^8$ \\
$\lambda_{\rm Edd}$              & \nodata &  \nodata                                            & \nodata                                          &  0.34 $\pm$ 0.20 \\
\enddata
\tablecomments{The velocity offsets were measured relative to \ion{Mg}{2} $\lambda$2800. The FWHMs were corrected for the instrumental velocity resolution.}
\end{deluxetable*}

\begin{figure}
\epsscale{1.2}
\plotone{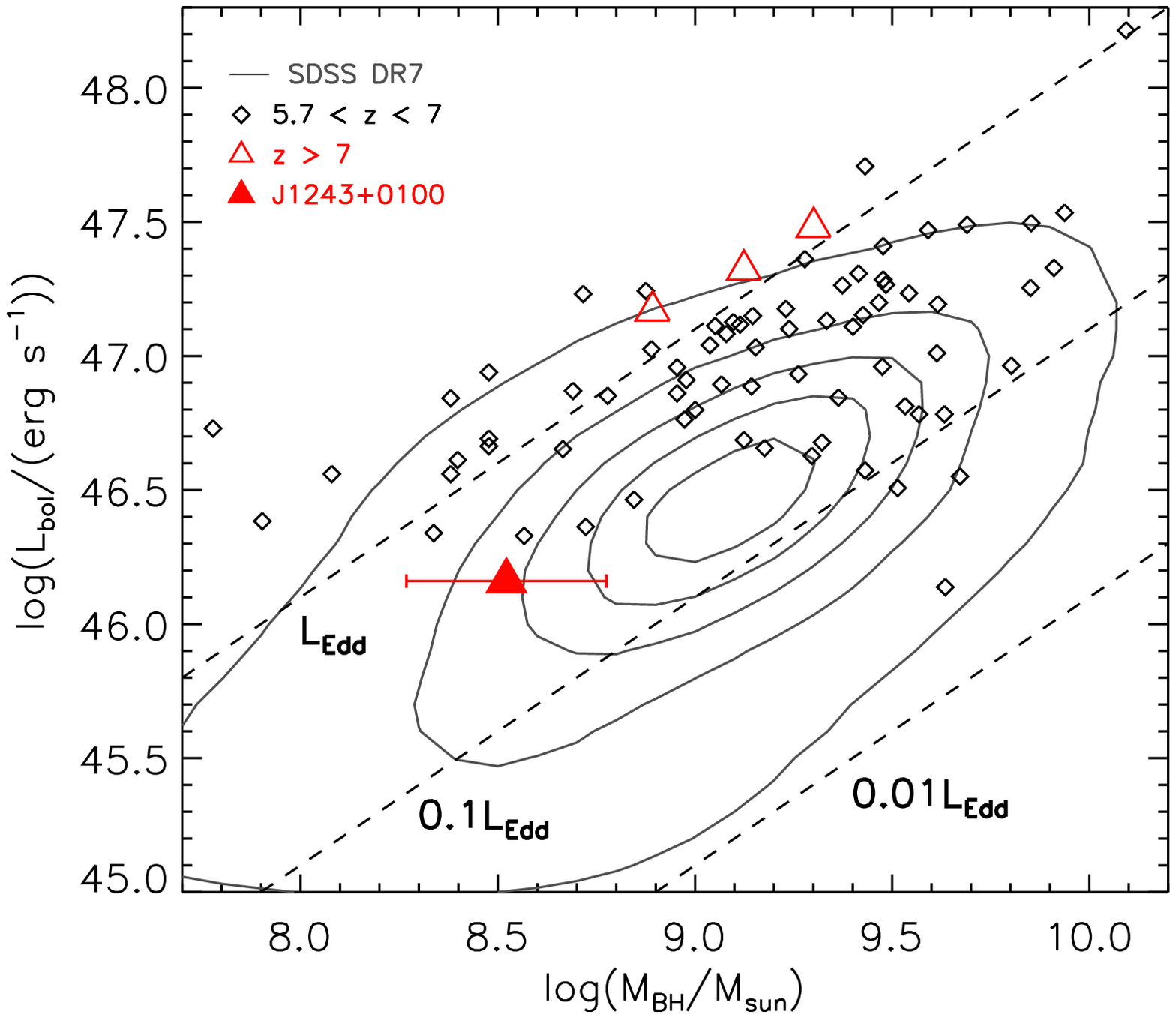}
\caption{Compilation of black hole mass and bolometric luminosity measurements in quasars.
The contours (linearly spaced in surface density) represent the distribution of quasars in the SDSS DR7 catalog \citep{shen11}, 
while the diamonds represent quasars at $5.7 < z < 7$ 
\citep{kurk07, jiang07, willott10, derosa11, derosa14, venemans15, wu15, mazzucchelli17, eilers18, shen18}.
The filled triangle represents $J1243+0100$, the quasar presented in this letter, while the unfilled triangles represent other $z > 7$ quasars reported by \citet{mortlock11}, \citet{banados18}, and \citet{wang18}.
\label{fig:Mbh_Lbol}}
\end{figure}

Figure \ref{fig:Mbh_Lbol} compares the estimated black hole mass and bolometric luminosity of $J1243+0100$ with those of other quasars in the literature. 
While the other known $z > 7$ quasars have $M_{\rm BH} \ga 10^9 M_\odot$ and radiate at the rates close to the Eddington limit, $J1243+0100$ has a considerably lower-mass black hole 
and is shining at a sub-Eddington rate. 
The luminosity and black hole mass of $J1243+0100$ are comparable to, or even lower than, those measured for the majority of low-$z$ quasars in the Sloan Digital Sky Survey 
(SDSS) Data Release 7 (DR7) catalog \citep{shen11}.
Thus, this quasar likely represents the first example of an ordinary quasar beyond $z = 7$.

On the other hand, given the limited S/N of the spectrum, the present measurements of $M_{\rm BH}$ and $\lambda_{\rm Edd}$ could be biased; 
for example, noise spikes can affect measurement of the \ion{Mg}{2} line width significantly.
We added artificial errors to the spectrum, based on the observed noise array, and performed the model fitting for 100 realizations of 
the generated spectrum.
This yielded $M_{\rm BH} =  (3.1 \pm 1.3) \times 10^8 M_\odot$ and $\lambda_{\rm Edd} = 0.36 \pm 0.16$ (the median values and intervals containing 
68\% of the realizations), while 5 and 2 \% of the realizations gave ($M_{\rm BH} < 10^8 M_\odot$, $\lambda_{\rm Edd} \ge 1$) and
($M_{\rm BH} > 10^9 M_\odot$, $\lambda_{\rm Edd} < 0.1$), respectively.
Deeper observations than presented here are clearly needed to better characterize this quasar.

$J1243+0100$ has strong emission lines with high equivalent widths, compared to the luminous $z > 7$ quasars, which may in part reflect the so-called \citet{baldwin77} effect.
We found that the rest-frame equivalent widths (REWs) of the emission lines  listed in Table \ref{tab:spec} are comparable to those of low-$z$ counterparts.
In particular, the median REWs of $\sim$13,000 SDSS DR7 quasars,
selected to have continuum luminosities within $\pm$0.1 dex of $J1243+0100$, are 49 \AA\ and 36 \AA\ for \ion{C}{4} and \ion{Mg}{2}, respectively.
This may indicate that the physical conditions in the broad line regions of quasars are similar from $z > 7$ to the relatively nearby universe.
In addition, 
the spectrum of $J1243+0100$ may show signs of BALs bluewards of the \ion{Si}{4} and \ion{C}{4} emission lines.
While these features are found at wavelengths relatively free from atmospheric absorption, the limited S/N of the present data prevents us from robustly confirming them.
This possible BAL signature and the large blueshift of the \ion{C}{4} emission line may indicate the presence of a fast gas outflow close to the quasar nucleus.

An important application of a high-$z$ quasar spectrum is to measure the IGM neutral fraction around the quasar, via the absorption damping wing analysis.
One could also estimate the quasar lifetime from an analysis of the quasar proximity effect.
However, such measurements require accurate modeling of the spectral shape around Ly $\alpha$, which is hard to do with the limited S/N of the present data.
The BAL features, if confirmed to be present, may also complicate such analyses for $J1243+0100$.
But these will be interesting subjects of follow-up studies, with deeper spectroscopy in the optical and near-IR.
Finally, future observations of this highest-$z$ ordinary quasar with, e.g., the Atacama Large Millimeter/ submillimeter Array and the {\it James Webb Space Telescope} 
will allow us to investigate the gaseous and stellar properties of the host galaxy,
which will be a key to understand the relationship between the quasar activity and the host galaxy at an early stage of cosmic history.

\section{Summary \label{sec:summary}}

This letter is the seventh in a series of publications presenting the results from the SHELLQs project, a search for low-luminosity quasars at $z \ga 6$ based on
the deep multi-band imaging data produced by the HSC-SSP survey.
We presented the discovery of $J1243+0100$, a quasar at $z = 7.07$.
It was selected as a quasar candidate from the HSC data, and its optical to near-IR spectrum was obtained with FOCAS and MOIRCS on Subaru, and GNIRS on Gemini.
The quasar has an order of magnitude lower luminosity than do other known quasars at $z > 7$.
The estimated black hole mass is $M_{\rm BH} = (3.3 \pm 2.0) \times 10^8 M_\odot$, and the Eddington ratio is 
$\lambda_{\rm Edd} = 0.34 \pm 0.20$.
As such, this quasar may represent the first example of an ordinary quasar beyond $z = 7$.
The large blueshift of the \ion{C}{4} emission line and possible BAL features suggest the presence of a fast gas outflow close to the quasar nucleus.

The discovery of $J1243+0100$ demonstrates the power of the HSC-SSP survey to explore SMBHs at $z > 7$, with masses typical of lower-$z$ quasars.
The quasar was selected from $\sim$900 deg$^2$ of the survey (including substantial area with partial survey depths), and we are in the course of follow-up observations of the remaining candidates. 
We expect to find a few more quasars at $z > 7$ by the completion of the survey, which is going to cover 1400 deg$^2$ in the Wide layer.
Combined with luminous $z > 7$ quasars discovered by other surveys, and also with lower-$z$ counterparts of ordinary quasars, those high-$z$ low-luminosity quasars will provide 
a significant insight into the formation and evolution of SMBHs across cosmic history.


\acknowledgments

This work is based on data collected at the Subaru Telescope and retrieved from the HSC data archive system, which is operated by the Subaru Telescope and Astronomy Data Center at 
National Astronomical Observatory of Japan (NAOJ).
The data analysis was in part carried out on the open use data analysis computer system at the Astronomy Data Center of NAOJ.

This work is also based on observations obtained at the Gemini Observatory and processed using the Gemini IRAF package.
The Observatory is operated by the Association of Universities for Research in Astronomy, Inc., under a cooperative agreement with the NSF on behalf of the Gemini partnership: the National Science Foundation (United States), the National Research Council (Canada), CONICYT (Chile), Ministerio de Ciencia, Tecnolog\'{i}a e Innovaci\'{o}n Productiva (Argentina), and Minist\'{e}rio da Ci\^{e}ncia, Tecnologia e Inova\c{c}\~{a}o (Brazil).

YM was supported by the Japan Society for the Promotion of Science (JSPS) KAKENHI Grant No. JP17H04830 and the Mitsubishi Foundation Grant No. 30140.
NK acknowledges supports from the JSPS grant 15H03645.
KI acknowledges support by the Spanish MINECO under grant AYA2016-76012-C3-1-P and MDM-2014-0369 of ICCUB (Unidad de Excelencia `Mar\'ia deMaeztu').

The Hyper Suprime-Cam (HSC) collaboration includes the astronomical communities of Japan and Taiwan, and Princeton University.  The HSC instrumentation and software were developed by the National Astronomical Observatory of Japan (NAOJ), the Kavli Institute for the Physics and Mathematics of the Universe (Kavli IPMU), the University of Tokyo, the High Energy Accelerator Research Organization (KEK), the Academia Sinica Institute for Astronomy and Astrophysics in Taiwan (ASIAA), and Princeton University.  Funding was contributed by the FIRST program from Japanese Cabinet Office, the Ministry of Education, Culture, Sports, Science and Technology (MEXT), the Japan Society for the Promotion of Science (JSPS),  Japan Science and Technology Agency  (JST),  the Toray Science  Foundation, NAOJ, Kavli IPMU, KEK, ASIAA,  and Princeton University.

The Pan-STARRS1 Surveys (PS1) have been made possible through contributions of the Institute for Astronomy, the University of Hawaii, the Pan-STARRS Project Office, the Max-Planck Society and its participating institutes, the Max Planck Institute for Astronomy, Heidelberg and the Max Planck Institute for Extraterrestrial Physics, Garching, The Johns Hopkins University, Durham University, the University of Edinburgh, Queen's University Belfast, the Harvard-Smithsonian Center for Astrophysics, the Las Cumbres Observatory Global Telescope Network Incorporated, the National Central University of Taiwan, the Space Telescope Science Institute, the National Aeronautics and Space Administration under Grant No. NNX08AR22G issued through the Planetary Science Division of the NASA Science Mission Directorate, the National Science Foundation under Grant No. AST-1238877, the University of Maryland, and Eotvos Lorand University (ELTE).
 
This letter makes use of software developed for the Large Synoptic Survey Telescope. We thank the LSST Project for making their code available as free software at http://dm.lsst.org.

IRAF is distributed by the National Optical Astronomy Observatory, which is operated by the Association of Universities for Research in Astronomy (AURA) under a cooperative agreement 
with the National Science Foundation.


\begin{thebibliography}{}
\bibitem[Aihara et al.(2018a)]{aihara17_survey} Aihara, H., Arimoto, N., Armstrong, R., et al.\ 2018, \pasj, 70, S4 
\bibitem[Aihara et al.(2018b)]{aihara17_pdr1} Aihara, H., Armstrong, R., Bickerton, S., et al.\ 2018, \pasj, 70, S8 
\bibitem[Baldwin(1977)]{baldwin77} Baldwin, J.~A.\ 1977, \apj, 214, 679 
\bibitem[Ba{\~n}ados et al.(2018)]{banados18} Ba{\~n}ados, E., Venemans, B.~P., Mazzucchelli, C., et al.\ 2018, \nat, 553, 473 
\bibitem[Chambers et al.(2016)]{chambers16} Chambers, K.~C., Magnier, E.~A., Metcalfe, N., et al.\ 2016, arXiv:1612.05560 
\bibitem[Decarli et al.(2017)]{decarli17} Decarli, R., Walter, F., Venemans, B.~P., et al.\ 2017, \nat, 545, 457 
\bibitem[De Rosa et al.(2011)]{derosa11} De Rosa, G., Decarli, R., Walter, F., et al.\ 2011, \apj, 739, 56 
\bibitem[De Rosa et al.(2014)]{derosa14} De Rosa, G., Venemans, B.~P., Decarli, R., et al.\ 2014, \apj, 790, 145 
\bibitem[Dey et al.(2018)]{dey18} Dey, A., Schlegel, D.~J., Lang, D., et al.\ 2018, arXiv:1804.08657 
\bibitem[Eilers et al.(2018)]{eilers18} Eilers, A.-C., Hennawi, J.~F., \& Davies, F.~B.\ 2018, \apj, 867, 30 
\bibitem[Elias et al.(2006)]{elias06} Elias, J.~H., Joyce, R.~R., Liang, M., et al.\ 2006, \procspie, 6269, 62694C 
\bibitem[Fan et al.(2006a)]{fan06araa} Fan, X., Carilli, C.~L., \& Keating, B.\ 2006, \araa, 44, 415 
\bibitem[Fan et al.(2018)]{fan18} Fan, X., Wang, F., Yang, J., et al.\ 2018, arXiv:1810.11924 
\bibitem[Ferrara et al.(2014)]{ferrara14} Ferrara, A., Salvadori, S., Yue, B., \& Schleicher, D.\ 2014, \mnras, 443, 2410 
\bibitem[Goto et al.(2009)]{goto09} Goto, T., Utsumi, Y., Furusawa, H., Miyazaki, S., \& Komiyama, Y.\ 2009, \mnras, 400, 843 
\bibitem[Gunn \& Peterson(1965)]{gunn65} Gunn, J.~E., \& Peterson, B.~A.\ 1965, \apj, 142, 1633 
\bibitem[Ichikawa et al.(2006)]{ichikawa06} Ichikawa, T., Suzuki, R., Tokoku, C., et al.\ 2006, \procspie, 6269, 626916 
\bibitem[Izumi et al.(2018)]{izumi18} Izumi, T., Onoue, M., Shirakata, H., et al.\ 2018, \pasj, 70, 36 
\bibitem[Jiang et al.(2007)]{jiang07} Jiang, L., Fan, X., Vestergaard, M., et al.\ 2007, \aj, 134, 1150 
\bibitem[Kashikawa et al.(2002)]{kashikawa02} Kashikawa, N., Aoki, K., Asai, R., et al.\ 2002, \pasj, 54, 819 
\bibitem[Kurk et al.(2007)]{kurk07} Kurk, J.~D., Walter, F., Fan, X., et al.\ 2007, \apj, 669, 32 
\bibitem[Lawrence et al.(2007)]{lawrence07} Lawrence, A., Warren, S.~J., Almaini, O., et al.\ 2007, \mnras, 379, 1599 
\bibitem[Lord (1992)]{lord92} Lord, S.D. 1992, NASA Technical Memor. 103957
\bibitem[Madau et al.(2014)]{madau14} Madau, P., Haardt, F., \& Dotti, M.\ 2014, \apjl, 784, L38 
\bibitem[Matsuoka et al.(2016)]{matsuoka16} Matsuoka, Y., Onoue, M., Kashikawa, N., et al.\ 2016, \apj, 828, 26
\bibitem[Matsuoka et al.(2018a)]{matsuoka18a} Matsuoka, Y., Iwasawa, K., Onoue, M., et al.\ 2018a, \apjs, 237, 5
\bibitem[Matsuoka et al.(2018b)]{matsuoka18b} Matsuoka, Y., Onoue, M., Kashikawa, N., et al.\ 2018b, \pasj, 70, S35
\bibitem[Matsuoka et al.(2018c)]{matsuoka18c} Matsuoka, Y., Strauss, M.~A., Kashikawa, N., et al.\ 2018c, \apj, 869, 150 
\bibitem[Mazzucchelli et al.(2017)]{mazzucchelli17} Mazzucchelli, C., Ba{\~n}ados, E., Venemans, B.~P., et al.\ 2017, \apj, 849, 91 
\bibitem[Mortlock et al.(2011)]{mortlock11} Mortlock, D.~J., Warren, S.~J., Venemans, B.~P., et al.\ 2011, \nat, 474, 616 
\bibitem[Oke \& Gunn(1983)]{oke83} Oke, J.~B., \& Gunn, J.~E.\ 1983, \apj, 266, 713 
\bibitem[Pacucci \& Loeb(2018)]{pacucci18} Pacucci, F., \& Loeb, A.\ 2018, arXiv:1810.12302 
\bibitem[Schlegel et al.(1998)]{schlegel98} Schlegel, D.~J., Finkbeiner, D.~P., \& Davis, M.\ 1998, \apj, 500, 525 
\bibitem[Shen et al.(2018)]{shen18} Shen, Y., Wu, J., Jiang, L., et al.\ 2018, arXiv:1809.05584 
\bibitem[Shen et al.(2011)]{shen11} Shen, Y., Richards, G.~T., Strauss, M.~A., et al.\ 2011, \apjs, 194, 45 
\bibitem[DES Collaboration(2005)]{des05} The Dark Energy Survey Collaboration 2005, arXiv:astro-ph/0510346 
\bibitem[Tsuzuki et al.(2006)]{tsuzuki06} Tsuzuki, Y., Kawara, K., Yoshii, Y., et al.\ 2006, \apj, 650, 57 
\bibitem[Vanden Berk et al.(2001)]{vandenberk01} Vanden Berk, D.~E., Richards, G.~T., Bauer, A., et al.\ 2001, \aj, 122, 549 
\bibitem[Venemans et al.(2015)]{venemans15} Venemans, B.~P., Ba{\~n}ados, E., Decarli, R., et al.\ 2015, \apjl, 801, L11 
\bibitem[Vestergaard \& Osmer(2009)]{vestergaard09} Vestergaard, M., \& Osmer, P.~S.\ 2009, \apj, 699, 800 
\bibitem[Vestergaard \& Wilkes(2001)]{vestergaard01} Vestergaard, M., \& Wilkes, B.~J.\ 2001, \apjs, 134, 1 
\bibitem[Volonteri(2012)]{volonteri12} Volonteri, M.\ 2012, Science, 337, 544 
\bibitem[Wang et al.(2018)]{wang18} Wang, F., Yang, J., Fan, X., et al.\ 2018, \apjl, 869, L9 
\bibitem[Willott et al.(2010)]{willott10} Willott, C.~J., Albert, L., Arzoumanian, D., et al.\ 2010, \aj, 140, 546 
\bibitem[Wright et al.(2010)]{wright10} Wright, E.~L., Eisenhardt, P.~R.~M., Mainzer, A.~K., et al.\ 2010, \aj, 140, 1868 
\bibitem[Wu et al.(2015)]{wu15} Wu, X.-B., Wang, F., Fan, X., et al.\ 2015, \nat, 518, 512 
\bibitem[Yang et al.(2018)]{yang18} Yang, J., Wang, F., Fan, X., et al.\ 2018, arXiv:1811.11915 
\end{thebibliography}
\end{document}